\documentclass{emulateapj}

\usepackage{rotating}

\newcommand{\til}{$\sim$}
\newcommand{\ergsqcmsec}{\thinspace\hbox{$\hbox{erg}\thinspace\hbox{cm}^{-2}
                \thinspace\hbox{s}^{-1}$}}
\newcommand{\ergsec}{\thinspace\hbox{$\hbox{erg}\thinspace\hbox{s}^{-1}$}}
\def\spose#1{\hbox to 0pt{#1\hss}}
\def\simlt{\mathrel{\spose{\lower 3pt\hbox{$\mathchar"218$}}
     \raise 2.0pt\hbox{$\mathchar"13C$}}}
\def\simgt{\mathrel{\spose{\lower 3pt\hbox{$\mathchar"218$}}
     \raise 2.0pt\hbox{$\mathchar"13E$}}}

\newcommand{\msun}{\thinspace\hbox{$M_{\odot}$}}
\newcommand{\cxo}{{\em Chandra}}
\newcommand{\rosat}{{\em ROSAT}}
\newcommand{\ginga}{{\em Ginga}}

\newcommand{\targ}{{V426~Oph}}

\def\today{\ifcase\month\or
January\or February\or March\or April\or May\or June\or
July\or August\or September\or October\or November\or December\fi
\space\number\day, \number\year}

\slugcomment{Accepted for publication in the Astrophysical Journal, \today}

\shorttitle{V426 Oph}
\shortauthors{Homer et al.}

\begin{document}


\title{
{\em Chandra} Observation of V426 Oph: Weighing the Evidence for a Magnetic White Dwarf}

\author{Lee Homer and Paula Szkody}
\affil{Astronomy Department, Box 351580, University of Washington, Seattle, WA 98195-1580}

\email{homer,szkody@astro.washington.edu}
\author{John C. Raymond} 
\affil{Harvard-Smithsonian Center for Astrophysics, 60 Garden Street,
Cambridge MA 02138}
\email{raymond@cfa.harvard.edu}
\author{Robert E. Fried\altaffilmark{1}}
\affil{Braeside Observatory, P. O. Box 906, Flagstaff, AZ 86002}
\author{D.W. Hoard}
 \affil{SIRTF Science Center, California Institute of Technology, 1200 E. California Blvd, Pasadena, CA 91125}
\email{hoard@ipac.caltech.edu}
\author{S. L. Hawley, M. A. Wolfe, J. N. Tramposch and K. T. Yirak}
\affil{Astronomy Department, Box 351580, University of Washington, Seattle, WA 98195-1580}
\email{slh@astro.washington.edu}\email{maw2323,jonica,ktyirak@u.washington.edu}
\altaffiltext{1}{Deceased}




\begin{abstract}
We report the results of a 45 ks \cxo\ observation of the cataclysmic variable V426 Ophiuchus. The high resolution
spectrum from the high-energy transmission grating spectrometer is most consistent with a cooling flow model, placing \targ\ among the group of CVs including U Gem and EX Hya.
An uninterrupted lightcurve was also constructed, in which we detect a significant 4.2~hr modulation together with its first harmonic at 2.1~hrs.  Reanalysis of archival \ginga\, and
\rosat\ X-ray lightcurves also reveals modulations at periods consistent with 4.2 and/or 2.1~hrs. Furthermore, optical photometry in V, simultaneous with the \cxo\ observation,
indicates a modulation anti-correlated with the X-ray, and later more extensive R band photometry finds a signal at \til2.1~hrs.  The earlier reported X-ray periods at \til0.5
and 1~hrs appear to be only transient and quasi-periodic in nature. 
In contrast, the 4.2~hr period or its harmonic are stable and persistent in X-ray/optical data from
1988 to 2003.  This periodicity is clearly distinct from the 6.85~hr orbit, and could be
due to the spin of the white dwarf. If this is the case, V426 Oph would be
the first long period intermediate polar with a ratio  $P_{spin}/P_{orb}$ of 0.6.
However, this interpretation requires unreasonable values of magnetic field
strength and mass accretion rate.
\end{abstract}


\section{Introduction}

The cataclysmic variable V426 Oph has proven to be somewhat enigmatic,
even though it is a bright (V=11.5--13.4) system that has been studied for a
number of years since its discovery as a nova-like with an emission line
spectrum \citep{herb60}. A detailed spectroscopic study by Hessman (1988)
revealed a K3 dwarf secondary in a 6.85~hr 
orbit with a white dwarf, with an inclination of 
$53^\circ$ and at a distance of 200 pc. He concluded this was a Z Cam type of dwarf
nova with outbursts approximately every 22 days and some 
standstills in the light
curve occurring about one magnitude fainter than the outburst brightness.
This interpretation would indicate V426 Oph is a system with such a high
rate of mass transfer that it is very close to the upper limit for dwarf novae
outbursts \citep{osak96}.

However, there are several observational clues that V426 Oph might also harbor
a magnetic white dwarf. IUE observations show an unusually
flat flux distribution in the UV, possibly indicating the truncation of a hot inner
disk by a magnetic white dwarf (i.e. an intermediate polar),
 while the X-ray flux at quiescence is larger than for typical dwarf novae
\citep{szko86}. The emission lines show flaring activity and a phase shift
from a location near the white dwarf \citep{hess88}. Most importantly, 
quasi-periodic variability has been seen in the optical and X-ray at a
period near 30 min at a brightness comparable to standstill \citep{szko90} and near 1, 2.5 or 4.5~hrs in the X-ray at quiescent 
brightness (Szkody 1986, Rosen et al. 1994; but see Hellier et al. 1990). 
While none of these periods could
be positively identified as that of the rotation of the white dwarf to confirm the identification as an intermediate polar, the different brightness states
combined with the lack of a long stretch of continuous observation 
have hampered attempts to resolve the origins of the observed periodicities.

With the advent of {\it Chandra}, high resolution X-ray spectroscopy can be
combined with uninterrupted long light curves to obtain much better
information on the accreting regions of cataclysmic variables. Results are
now available for the dwarf novae U Gem \citep{szko02}, and  WX Hyi \citep{pern03},
the intermediate polar EX Hya \citep{mauc01}, along with 5 other cataclysmic variables \citep{muka03}.
One of the surprising results is how similar the spectrum of the
 low mass accretion disk
system U Gem is to that of the intermediate polar EX Hya. \citet{muka03}
have found that the seven systems can be divided into two groups. The first group,
which includes U Gem and EX Hya, shows a bremsstrahlung continuum with strong
H and He-like ion emission as well as ions from FeXVII-FeXXIV. The other
group shows a harder X-ray continuum and little \linebreak Fe L-shell emission.
They suggest the differences in these two groups may be due to differences
in the accretion rate per unit area. In the first group, the energy is
released as optically thin radiation in a steady state cooling flow. In
the second, the line emission arises from a plasma photoionized by the hard
continuum. To further explore these ideas, and to try to resolve the nature
of the accretion in V426 Oph, we obtained a {\it Chandra} observation at
 quiescence, 
together with  
 simultaneous optical photometry and
a single optical spectrum. We also obtained additional optical photometry (in quiescence) at a later date.

\section{Observations}
\begin{deluxetable*}{lccc}
\tablewidth{0pt}
\tablecaption{Observation Summary\label{tab:obslog}
}
\tablehead{
\colhead{UT Date} & \colhead{Obs} & \colhead{UT Time} &
\colhead{Comments} }
\startdata
2002 May 30, 31 & Chandra: ACIS-S/HETGS & 21:40 -- 11:02 & 45.15 ks good time \\
2002 May 31 & BO & 07:04 -- 11:30 & V filter photometry \\
2002 May 30& CTIO: YALO & 08:17 -- 09:34 & V filter photometry \\  
2002 May 31 & APO: DIS &  06:20 -- 06:25 & spectrum \\
2003 July 25 & MRO & 06:00 -- 10:04& R filter photometry\\
2003 July 28 & MRO & 04:35 -- 10:51& I filter photometry\\
\enddata
\end{deluxetable*}

The {\it Chandra} observation of V426 Oph started at 21:40 UT on 2002 May 30
and ended at 11:02 UT the following day, with 45.15 ks of good exposure time
on the source.
  
Differential photometry with respect to comparison star 1 in \citet{miss96}
 was accomplished with the
Braeside Observatory (BO) 0.4m reflector using a SITe 512 CCD camera and a 
Bessell V filter on 
2002 May 31 UT. 
Fifty second integrations for 4.4~hrs provided simultaneous overlap of 3.5~hrs
with the Chandra data. Additional contemporaneous photometry was taken by the YALO 1m telescope at Cerro-Tololo Inter-American Observatory
 (CTIO). Using comparison stars 13, 14 and 15 \citep{hend97}, the V magnitude of V426 Oph was determined to
be $12.79\pm0.15$ (YALO) and using star 1,  between 12.7-12.9 (BO), consistent
with a quiescent state. Observations on the AAVSO and VSNET web sites show
that the previous outburst (V$\sim$11) occurred on May 13-14.

On 2003 July 25 and 28 UT longer duration differential photometry in Harris R (4.0~hrs) and I (6.3~hrs) was obtained at Manastash Ridge Observatory (MRO), with its
0.75 m reflector. The exposure times were 30 and 10~s, respectively.

\begin{figure}[!h]
\resizebox{.46\textwidth}{!}{\rotatebox{0}{\plotone{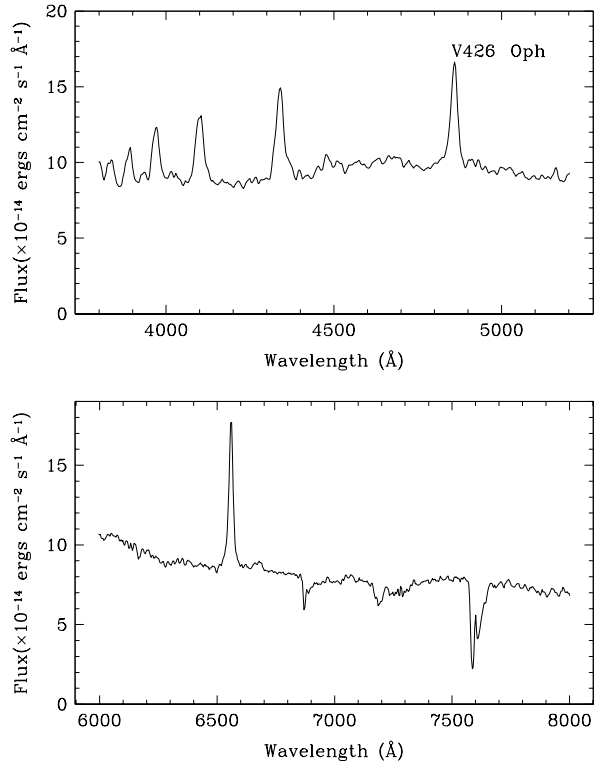}}}
\caption{Low-resolution spectrum (6\AA) of \targ\ 
obtained during the \cxo\ observation, note the broad Balmer emission lines. \label{fig:APOspec}}
\end{figure}

A single low resolution optical spectrum (6\AA) was also
obtained during the \cxo\ observation, using the Double Imaging Spectrograph
(DIS) on the 3.5m Apache Point Observatory (APO) telescope, providing
spectral coverage from 3800-5200\AA\ and 6000-8000\AA\ (Figure~\ref{fig:APOspec}). This
spectrum also confirms that V426 Oph was at a quiescent state, exhibiting 
characteristic broad Balmer emission lines. All the observations are summarized
 in  Table~\ref{tab:obslog}.
\vspace*{5mm}
\pagebreak

\section{ X-ray spectral analyses and results}
We reprocessed our data starting from the level 1.5 event file, as advised by the \cxo\ X-ray Center.  To check for intervals of high background
we constructed an off-source lightcurve, but as none were found we were able to use the standard good time intervals as supplied.  We chose to remove after-glow
detection, as for these data it may have led to exclusion of good events.  We applied destreaking to the S4 chip, to produce a clean level 2 event
file, at the same time restricting our energy range to 0.3--10.0 keV, where calibrations are most reliable.  We used
 {\tt tgextract} to derive the type II pulse height analyzer (PHA) file.  To achieve maximum signal-to-noise (S/N), we opted to add +1 and -1 grating orders, for both
the medium (MEG) and high (HEG) energy grating spectra, creating type I PHA files.  The script {\tt add\_grating\_orders} also generated  the appropriate grating response files for these co-added spectra.  For high-resolution
spectra it is generally advised to include the systematic sensitivity degradation of the ACIS chips (due to a build-up of absorbing material)
as an independent model component.  However we found that our S/N was
insufficient to constrain the individual elemental abundances of the material.  Instead, we corrected the ancillary response files  (ARF) for the expected absorption at the time of observation, based on the nominal elemental composition.
The HEG and MEG spectra are shown in Figure 2.
\pagebreak 
\subsection{Cooling flow fits}
First, in order to compare to earlier work we fitted a limited, mostly line free section of
our spectrum from 2.0 to 5.0 keV with a simple thermal bremsstrahlung plus absorption model.  Our best fit yielded a poorly constrained column of
$N_H=0.57\pm0.54\times10^{22}{\rm cm}^{-2}$, and $kT_{Br}=20\pm12$keV, with a reduced $\chi^2_{\nu}=0.51$ for 827 degrees of
freedom.  Both from this and simple inspection of the spectra we can see that \targ\ belongs to the first group of CVs identified by
\citet{muka03}; namely those best fit by a cooling flow as opposed to a photo-ionization model, with a maximum temperature nearer 20 than 80 keV.
 
We therefore proceeded to attempt a global cooling flow fit to our 
high-resolution spectra, using XSPEC\footnote{The X-ray software packages, XSPEC (spectral) and FTOOLS (general) are both available from http://heasarc.gsfc.nasa.gov/docs/software/lheasoft/}  with the same model as \citet{muka03}.
  A suitable background file was obtained with the CIAO\footnote{The Chandra data analysis software, CIAO, is available from http://asc.harvard.edu/ciao/} {\tt tg\_bkg} script.  We used the FTOOL\footnotemark[2] {\tt grppha}
to associate the appropriate redistribution matrix (RMF), ARF and background files with the co-added first
order grating PHA files.  We also used  {\tt grppha} to group the spectra both simply binning by 4, then with binning to give $\geq10$
counts per bin.  The S/N of our spectra is such that we opted to use the $\geq10$ counts per bin, and used $\chi^2$ statistics for fitting.

\begin{figure*}[!htb]
\resizebox{.85\textwidth}{!}{\rotatebox{0}{\plotone{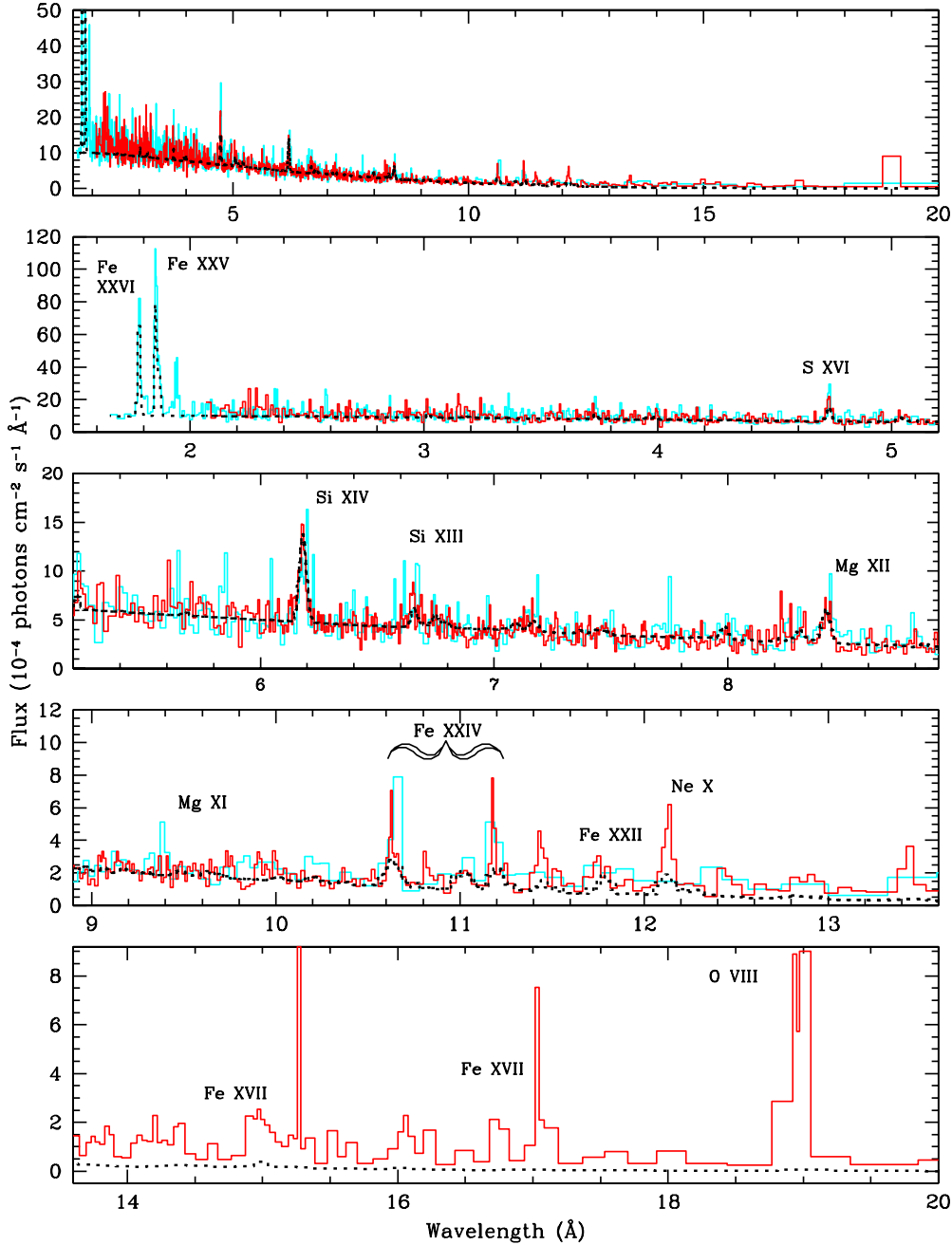}}}
\caption{\cxo/HETGS unfolded spectrum of \targ; both HEG (light grey, solid cyan in electronic edition) and MEG (dark grey, solid red in electronic edition) are shown.  Overlaid is the best fit cooling
  flow model (dotted black).  In the top panel we show an overview, while beneath we expand the wavelength scale and identify key emission
  lines.  Note: the lowest panel shows the spectrum binned with S/N$\geq$2 for clarity,
  the rest have $\geq$10 counts per bin. \label{fig:cxospec}}
\end{figure*}

In order to attain the best constraints possible we performed joint fits to 
both the HEG and MEG spectra, over wavelength ranges of 1.7--15~\AA\
and 2.2--20~\AA\ respectively, where counts are most adequate for fitting. Initially, we constrained the column to the $N_H=0.7$, {1.6} and
$0.28\times 10^{22}$cm$^{-2}$ found by {\it EXOSAT}, {\it Ginga} and {\it ROSAT} observations respectively \citep{szko86,szko90,rose94}.  The
intermediate column from the {\it EXOSAT} spectrum gave the best fit, with $\chi_{\nu}^2=0.55$. Indeed, with the column left free we achieved a best fit for $N_H=1.0\times 10^{22}$cm$^{-2}$, giving
a $\chi_{\nu}^2=0.47$ for 1718 d.o.f.  The parameters for  these four fits along with the corresponding 2--10 keV fluxes are summarized in
Table~\ref{tab:cffit}, and the best model fit is over-plotted 
in Figure~\ref{fig:cxospec}.  As found for WX Hyi \citep{pern03}, we still find that our best fit model fails to fit any of the prominent
lines at $\simgt$13\AA\, notably OVIII. Even setting low $T = 0.08$~keV, the minimum allowed in XSPEC, did not enhance these line fluxes noticeably. 

\begin{deluxetable*}{lcccccc}
\tablewidth{0pt}
\tablecaption{Critical Parameters for Cooling Flow Fits \label{tab:cffit}
}
\tablehead{
\colhead{$N_H$ } & \colhead{low T (keV)} & \colhead{high T (keV)}&\colhead{Line width @ 6 keV}&  \colhead{\.M ($M_{\odot}\; {\rm
	 yr}^{-1}$)}&\colhead{$\chi_{\nu}^2$}&\colhead{2--10 keV flux}\\
\colhead{($10^{22} {\rm cm}^{-2}$)} &&&\colhead{$\sigma_v$(km s$^{-1}$)}&&&\colhead{ ($10^{-11}$\ergsqcmsec)}}
\startdata

0.28\tablenotemark{a} & $0.9\pm1.0$ & $24.0\pm0.9$ &$600\pm100$& $9.6\pm0.8\times10^{-12}$ &0.95 &$1.8\pm0.2$\\
0.7\tablenotemark{a}&$1.1\pm1.0$&$24.2\pm0.9$&$500\pm100$&$1.3\pm0.2\times10^{-11}$& 0.55 & $2.6\pm0.4$\\
1.6\tablenotemark{a} & $1.1\pm2.4$ &$ 20.0\pm1.1$ &$600\pm400$& $1.9\pm0.5\times10^{-11}$& 0.60&$3.6\pm0.9$\\
$1.0\pm0.1$&$1.0\pm0.2$&$24.2\pm0.4$ & $600\pm100$& $1.6\pm0.2\times10^{-11}$& 0.47& $3.0\pm0.3$\\
\tablenotetext{a}{Parameter values without uncertainties are fixed.}
\enddata
\end{deluxetable*}

Hence, we also experimented with several variations on the cooling flow
model using a radiative shock wave code \citep{raym79}.  We chose a
shock speed of 4250~$\rm km~s^{-1}$, to give a desired temperature of 22 keV,
       and complete electron-ion thermal equilibration.  In this case, we compared the model results to specific line fluxes and ratios. IDL routines were used to extract background subtracted and binned data in wavelength and flux units for both HEG and MEG. To estimate line
fluxes for the stronger lines we used the IRAF\footnote{IRAF (Image Reduction and Analysis Facility) is distributed by the National Optical Astronomy Observatory,
  which is operated by the Association of Universities for Research in Astronomy (AURA) Inc., under cooperative agreement with the National
  Science Foundation} {\tt splot} task to sum the flux over the width of the line above a continuum estimated by hand.
For the weaker lines, important for line diagnostics (see next section), we instead used
Gaussian fits to short segments of the MEG spectrum using binning with S/N$\geq$2 per bin, constraining the line widths to be the same. The various line measurements are presented in Table~\ref{tab:linespec}.

 The
shock code allows us to examine the effects of photoionization.
We can also make the flow nearly isobaric or nearly isochoric
by adjusting the assumed magnetic field.  These experiments
showed that the line ratios are essentially independent of
whether the flow is isobaric or isochoric, as expected, except
that photoionization is relatively more important in the constant
density case.  Radiation from the cooling gas does affect the
ionization state at the lower temperatures, but even for
constant density cooling the effects are only 10-20\% reductions
in ratios such as O~VIII/Fe~XXV.  A comparison of line intensities
(Table~\ref{tab:linespec}) indicates that $N_H = 0.28 \times 10^{22}~\rm cm^{-2}$
best matches the O~VIII intensity, while $N_H = 0.57 \times 10^{22}~\rm cm^{-2}$
does better for Ne X, Mg XII and Fe~XVII.  Within the restriction
of the cooling flow model either value of $N_H$  provides an
adequate fit.
 
We also examined the effects of thermal conduction.  As found by
\citet{pern03} for the case of WX~Hyi, thermal conduction
transports energy to the cooler part of the flow, drastically
increasing the strength of lines such as O VIII and Ne X.  Even
saturated thermal conduction \citep{cowi77} yields
large ratios of low-T to high-T lines unless the conduction is strongly suppressed.

\subsection{Line diagnostics}

\begin{deluxetable*}{lccccc}
\tablewidth{0pt}
\tablecaption{Line Summary\label{tab:linespec}
}
\tablehead{
\colhead{Line} & \colhead{Rest $\lambda$} &  
\colhead{MEG Flux\tablenotemark{a}} & \colhead{HEG Flux\tablenotemark{a}} &
\multicolumn{2}{c}{Shock} \\
&\AA&&&$N_H = 0.28 \times 10^{22}~\rm cm^{-2}$&$N_H = 0.57 \times 10^{22}~\rm cm^{-2}$  }
\startdata
Fe XXVI & 1.783 &\nodata& 166  & 128 & 127\\
Fe XXV & 1.850 &\nodata&191   & 216  & 214 \\
S XVI  & 4.729 & 16  & 20  & 26.2& 24.6\\
Si XIV & 6.180, 6.187 & 13  & 11  & 30.2 &26.6\\
Si XIII & 6.647-6.688 & 6.8 & 9& 17.7&15.2\\
Mg XII & 8.419, 8.424 & 3.5 & 3.4 & 7.31& 5.52\\
Fe XXIV & 10.620-10.662& 3.1 & 8.7 & 23.2\tablenotemark{b} & 15.1\tablenotemark{b}\\
Fe XXIII & 11.014 &\nodata     &3.2  &8.27 & 5.10 \\
Fe XXIV & 11.172, 11.189 & 2.2 & 4.4   &27.0 & 14.8\\
Fe XXIV & 11.429 & 2.3 &1.5 &10.0 &5.33\\
Fe XXIII & 11.740 & 1.6&\nodata     &13.7&6.86\\
Fe XXII & 11.771 & 0.3  &\nodata     &6.41&3.20\\
Ne X & 12.128 & 3.1 &\nodata     & 12.4 &6.18 \\
Fe XVII & 15.012 & 4.2  &\nodata &7.5     &2.88   \\
Fe XVII & 16.774 & 2.8 &\nodata &3.35&0.73\\
Fe XVII & 17.051        & 3.4 &\nodata  &3.51&0.72\\
Fe XVII &        17.098 & $<$2.8&\nodata  &2.66&0.56\\
O VIII & 18.970 & 5.9 &\nodata & 17.7& 3.29\\
\enddata
\tablenotetext{a}{flux units are
10$^{-14}$\ergsqcmsec}
\tablenotetext{b}{Models are with the blend of Fe XXIV 10.620, 10.662}
\end{deluxetable*}

\citet{mauc01,mauc03} have shown that the ratios of various Fe ion species  can be a useful diagnostic for the density of the line
emitting plasma. Using our Gaussian line fits, we have calculated these ratios, which are presented alongside those of the Sun and EX Hya
\citep[from][]{mauc01} in
Table~\ref{tab:linerats}.   Unfortunately, with our low S/N,
the ratios for V426 Oph are only upper limits which are consistent with
EX Hya, and only differ with the Sun in the ratios involving the 17.10\AA\ line.
Thus, they do not provide useful diagnostics for the plasma densities nor
temperatures. 
\begin{deluxetable}{lccc}
\small
\tablewidth{0pt}
\tablecaption{Fe XVII and XXII line ratios\label{tab:linerats}
}
\tablehead{
\colhead{Ratio} & \colhead{Sun} &
\colhead{EX Hya} &
\colhead{V426 Oph}}
\startdata
11.92/11.77 & \nodata & 1.06 & $<2$\\
15.01/16.78 & 1.04 & 1.23 & $<3.9$ \\
15.26/16.78 & 0.51 & 0.50 & $<2.0$ \\
15.45/16.78 & \nodata & 0.05 & $<0.16$ \\
17.05/16.78 & 1.40 & 1.65 &$<1.8$ \\
17.10/16.78 & 1.32 & 0.08 & $<$1.0 \\
15.26/15.01 & 0.49 & 0.41 & $<$0.8 \\
17.10/17.05 & 0.93 & 0.05 & $<$0.8\\
\enddata
\end{deluxetable}
\section{Temporal analyses and results}
\begin{figure*}[!htb]
\resizebox{.95\textwidth}{!}{\rotatebox{-90}{\plotone{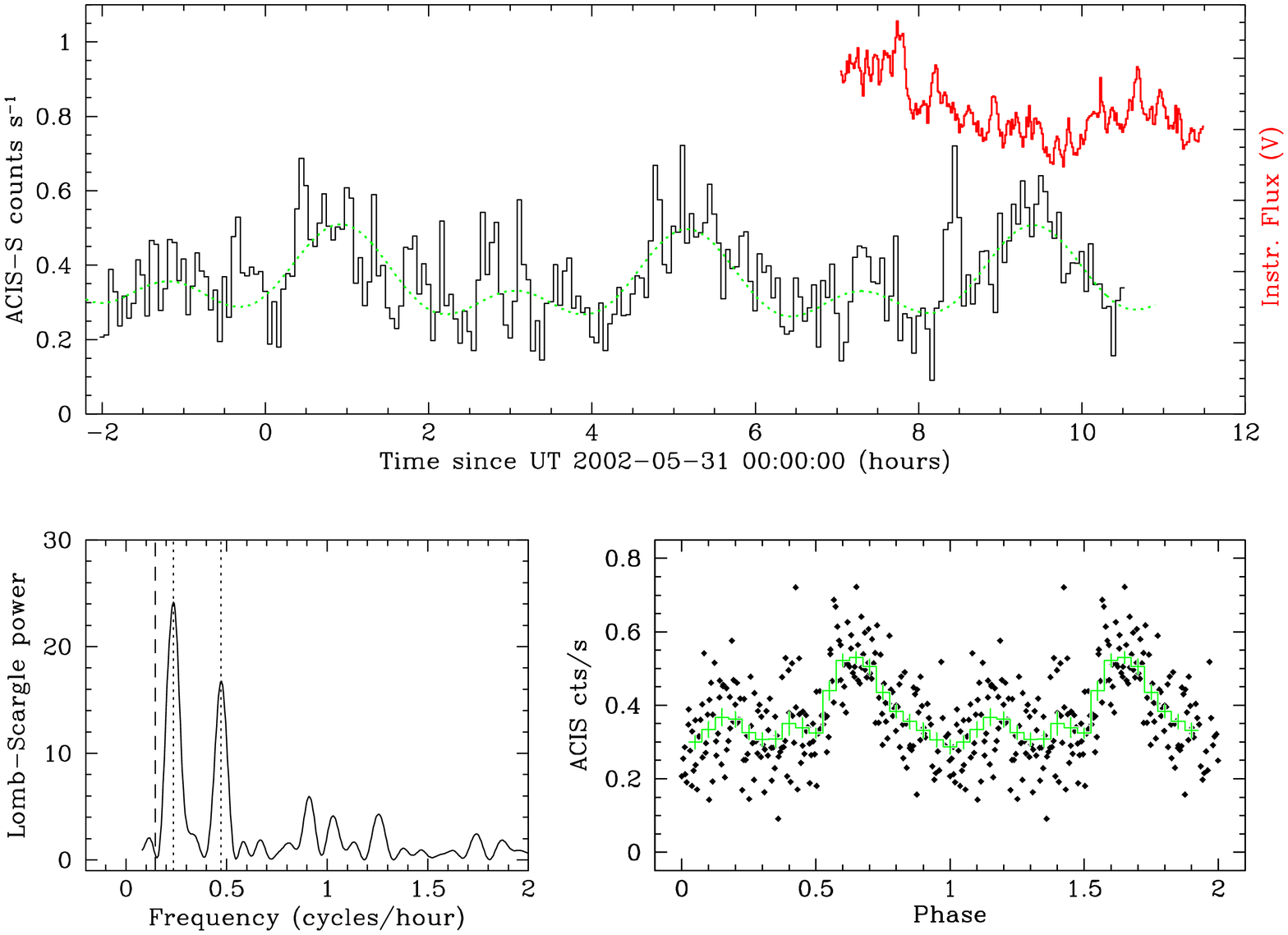}}}
\caption{Top panel: Partly simultaneous lightcurves:  lower, black curve-- X-rays from \cxo/ACIS-S, and upper, dark grey (red in electronic edition) curve-- optical(differential V band magnitudes converted to arbitrary
  count rate units for clarity). A double-sinusoid fit, with periods of 4.23 and
2.12~hrs, is over-plotted on the X-ray lightcurve.  Lower left: Lomb-Scargle periodogram of the ACIS data.  Two highly significant peaks  ($>99.99$\% confidence) are
  apparent; the vertical bars indicate (left to right) the radial velocity determination of the binary orbital period (6.85~hrs), the fundamental
  of a double-sinusoid fit to the lightcurve (4.2~hrs) and its first harmonic (2.1~hrs).  Lower right:  The ACIS lightcurve folded on 4.2~hrs, and
  overlaid with binned values.\label{fig:simlc}}
\end{figure*}

\subsection{\cxo\ plus archival {\em ROSAT} and  \ginga\ data}
The core of the \cxo\ 0th order image is significantly piled-up (\til50\%), hence this could not be used to construct a broad-band lightcurve.  Instead, an event
file was created from our reprocessed level 2 file, selecting only events in the 1.5 to 9.0~keV range, which corresponds to the peak of the
source spectrum, hence minimizing background contributions. Extraction apertures were chosen to include only the wings of the
0th order image PSF, plus the brightest parts of the 1st order spectrum landing on the S3 and S4 chips; the
areas close to chip boundaries were specifically excluded, since they have a reduced exposure fraction due to the spacecraft dither.
As a cross-check, a lower signal-to-noise lightcurve using solely the PSF wings was subjected to the same analysis; fully-consistent results were
found.  The CIAO {\tt lightcurve} tool was used to extract a lightcurve with 200s bins, with background subtraction
 based upon corresponding regions each side of the dispersed spectra; this is plotted in Figure~\ref{fig:simlc} (top panel).

To search for periodicities in the lightcurve, both a Lomb-Scargle \citep{scar82} and Phase-Dispersion Minimization \citep{stell78}
periodogram were calculated; the former is most sensitive to sinusoid 
modulations, while the latter is better for periodic signals with
more irregular morphology. Significant peaks ($\simgt99.99\%$ level) were found at two frequencies corresponding to 2.13 and 4.31 hours, as shown in
Figure~\ref{fig:simlc}, lower left.  We then directly fit the lightcurve with two sinusoids plus a
second order polynomial to take into account any longer term trends.  
The 2.1~hr period is close enough to being the 1st harmonic of the 4.3~hr that we
also constrained the periods to these values, leading to period
determinations of  $4.23\pm0.05$ and
$2.12\pm0.02$~hrs.  Indeed, these latter periods do produce a good fit to the data (see Fig.~\ref{fig:simlc}, top panel), with a reduced $\chi^2=3.03$ as compared to
$\chi^2_{\nu}=3.01$ for the unconstrained.    The total peak-to-peak amplitude 
is large at $65\pm5$\%. Interestingly these two periods are close to two of the 
candidate periods found by \citet{rose94} in their
{\em ROSAT} lightcurve. However, we note that we found no evidence for signals at either of the putative short periods, \til60~min or
\til30~min \citep{szko86,szko90}, the binary orbit at 6.85~hrs \citep{hess88} nor the 1.25~hr {\em ROSAT} candidate.  Neither did we find signals at
either of the spacecraft dither periods of 16.67 or
11.78~min, implying that we successfully excluded regions of reduced exposure 
fraction.

\begin{figure}[!htb]
\resizebox{.47\textwidth}{!}{\rotatebox{-90}{\plotone{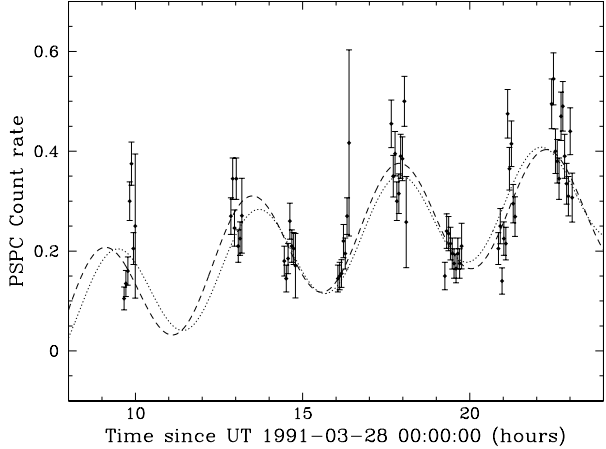}}}
\caption{X-ray (0.5--2.5~keV) lightcurve  from a {\em ROSAT}/PSPC pointed observation of V426 Oph.  Overlaid are sinusoid fits, both a free fit
  (dashed line) with a period of 4.43~hrs, and another (dotted) with the period fixed at that found from the \cxo\ data, 4.23~hrs.\label{fig:roslc}}
\end{figure}

To compare directly to the {\em ROSAT} data, we extracted the archival 200s binned PSPC lightcurve from the HEASARC.  The seven spectral bands
were co-added to give us maximal S/N.  We examined whether this data could also be fit by the periods we found in the high S/N uninterrupted \cxo\ lightcurve.
A free fit found a period of $4.43 \pm0.08$~hrs with $\chi^2_{\nu}=3.24$, consistent with that found by \citet{rose94}, while if we constrained the period to 4.23~hrs  the fit was not significantly
worse with $\chi^2_{\nu}=3.49$.  We also attempted a double-sinusoid fit, as for the morphology seen in the \cxo\ data; here the modulation
is in contrast best fit by a single sinusoid.  Both single sinusoid fits are shown overlaid on the data in Figure~\ref{fig:roslc}.

\begin{figure}[!htb]
\resizebox{.45\textwidth}{.5\textheight}{\rotatebox{0}{\plotone{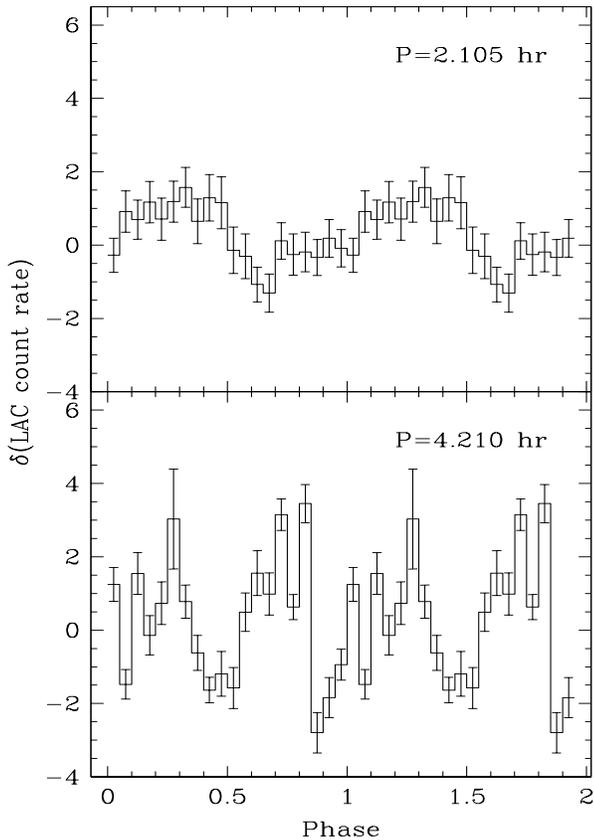}}}
\caption{X-ray (2--6~keV) lightcurve  from a {\em Ginga}/LAC pointed observation of V426 Oph.  The data have been folded on periods of 2.105~hrs
  and 4.21~hrs respectively, then binned for clarity. \label{fig:gingalc}}
\end{figure}

The {\it Ginga} data have a long time-base (72~hrs) but, like {\em ROSAT}, interrupted sampling due to Earth occultations. We obtained 64s
binned lightcurves from the HEASARC, for the 2--6~keV range (where the spectrum peaks and S/N is best), then applied a quadratic detrend to
remove longer term variations. The periodograms revealed evidence of significant flickering but
peaks close to the 4.2~hr and its harmonic were by far the most significant.  Once again a constrained double sinusoid fit yielded periods
of $2.105\pm0.005$~hrs and $4.210\pm0.010$~hrs, consistent with the \cxo\ data.  However, the fit has $\chi^2_{\nu}=15$ for the 863 d.o.f, probably as a result of the extensive
flickering, but also the morphology of the profile. In Fig.~\ref{fig:gingalc} we show folded and phase binned data, where the binning should
average out the effects of flickering.  Although the modulation is modulated primarily at the harmonic (2.1~hrs), the differing morphology of
the two humps in the lower fold accounts for the significance of the 4.2~hr signal.

\citet{hell90}
 reexamined the {\em EXOSAT} data in which \citet{szko86} detected a \til60 min periodicity and suggested an intermediate polar
classification for \targ.  They concluded that the modulation was not in fact statistically significant.  Moreover, they
noted that the modulation on the white
dwarf's spin should be greatest at low energies, since it is due to absorption effects.  Hence, the hardness ratio (hard/soft flux) should be
 anti-correlated with the total flux.
We have investigated the energy dependence of our significant 4.23~hr modulation.  We extracted 800~s binned lightcurves in two energy bands, 0.3--2.5~keV for the soft and 2.5--6.0~keV for the hard, which approximately splits the counts evenly.  In both bands the modulation is apparent, and indeed the
signals are significant at $\simgt$99.95\% confidence level.  Although the percentage amplitudes are different, with a hard band peak--peak amplitude of
$59\pm5$\% versus $73\pm6$\% in the soft, the hardness 
ratio values are consistent with a constant, the signal at 4.23~hrs
being merely 5\% confidence.  We also applied  a Spearman rank correlation test comparing the hardness ratio with the full-band lightcurve.
This yielded the expected anti-correlation though only at the 2$\sigma$ level. 

\subsection{Optical results}
We succeeded in obtaining optical photometry which overlaps the \cxo\ X-ray by 3.5 hours at the end of the observation.  This lightcurve is
plotted  for ease of comparison in Figure~\ref{fig:simlc} (top panel).  Although the simultaneous coverage is limited, it is clear that the
optical variability is in
general {\em anti}-correlated with the X-ray, and that it also exhibits an \til4.2~hr periodicity. A Spearman rank correlation test yields an anti-correlation at the 3.7$\sigma$ level.
\begin{figure*}[!htb]
\resizebox{.9\textwidth}{!}{\rotatebox{0}{\plotone{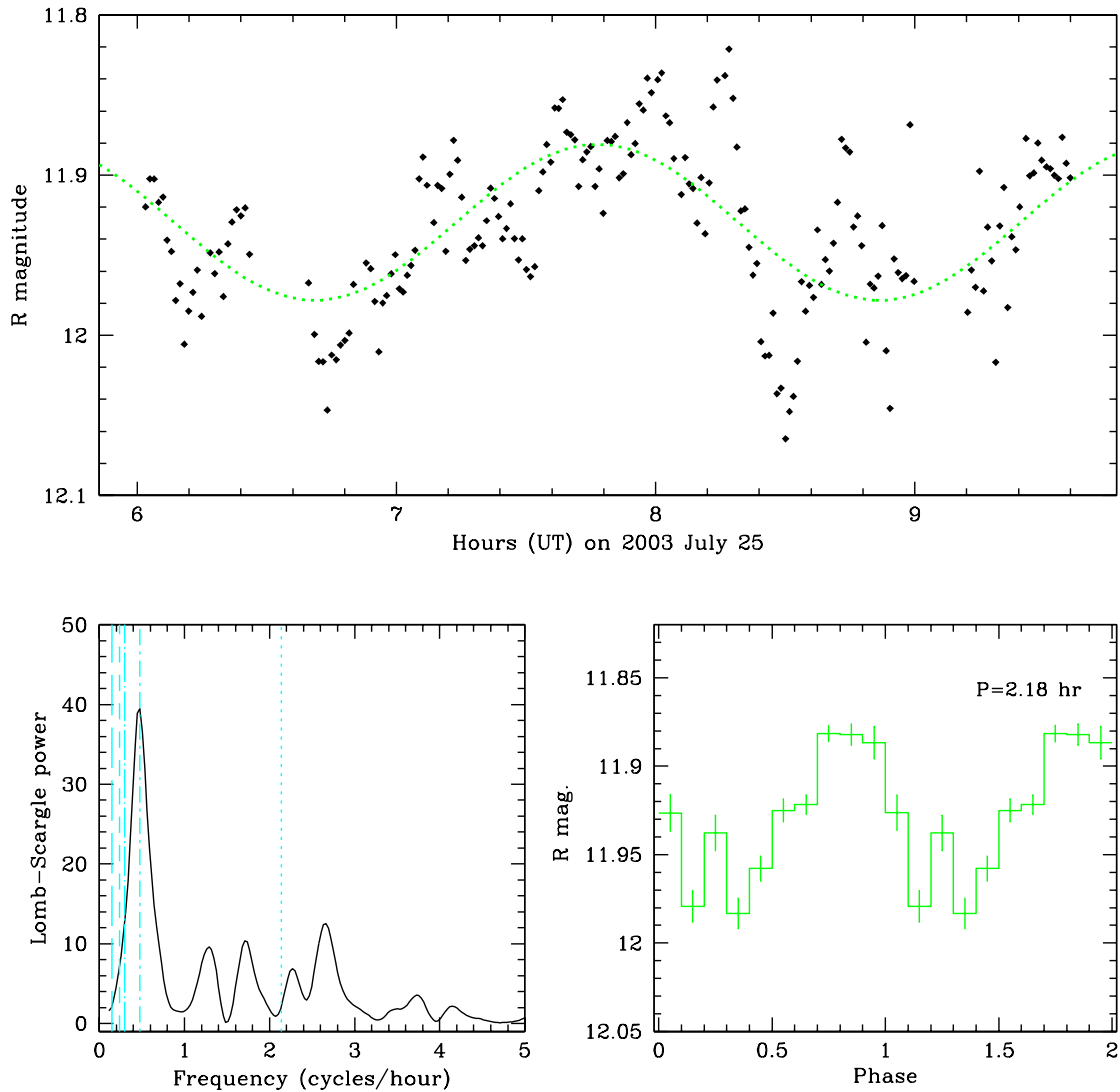}}}
\caption{Top panel: Differential $R$ band lightcurve of \targ, a sinusoid fit with  $P=2.18$~hrs is over-plotted. Lower left: Lomb-Scargle
  periodogram of $R$ data. A number of significant peaks  ($>99$\% confidence) are
  apparent, the largest corresponding to the 2.18~hr modulation, while the rest are likely due to the prominent flickering.  The vertical
  bars indicate (left to right) the radial velocity determination of the binary orbital period (6.85~hrs), the X-ray signal (4.2~hrs), first
  harmonic of the orbit (3.42~hrs), and the first harmonic of the X-ray signal (2.1~hrs), plus the 28 min period from \ginga.  Clearly, the $R$ band modulation is
  consistent with the
  first harmonic of the X-ray signal.  Lower right:  The $R$ band lightcurve folded on 2.18~hrs and binned.\label{fig:mro_Rlc}}
\end{figure*}
\begin{figure*}[!htb]
\resizebox{.9\textwidth}{!}{\rotatebox{0}{\plotone{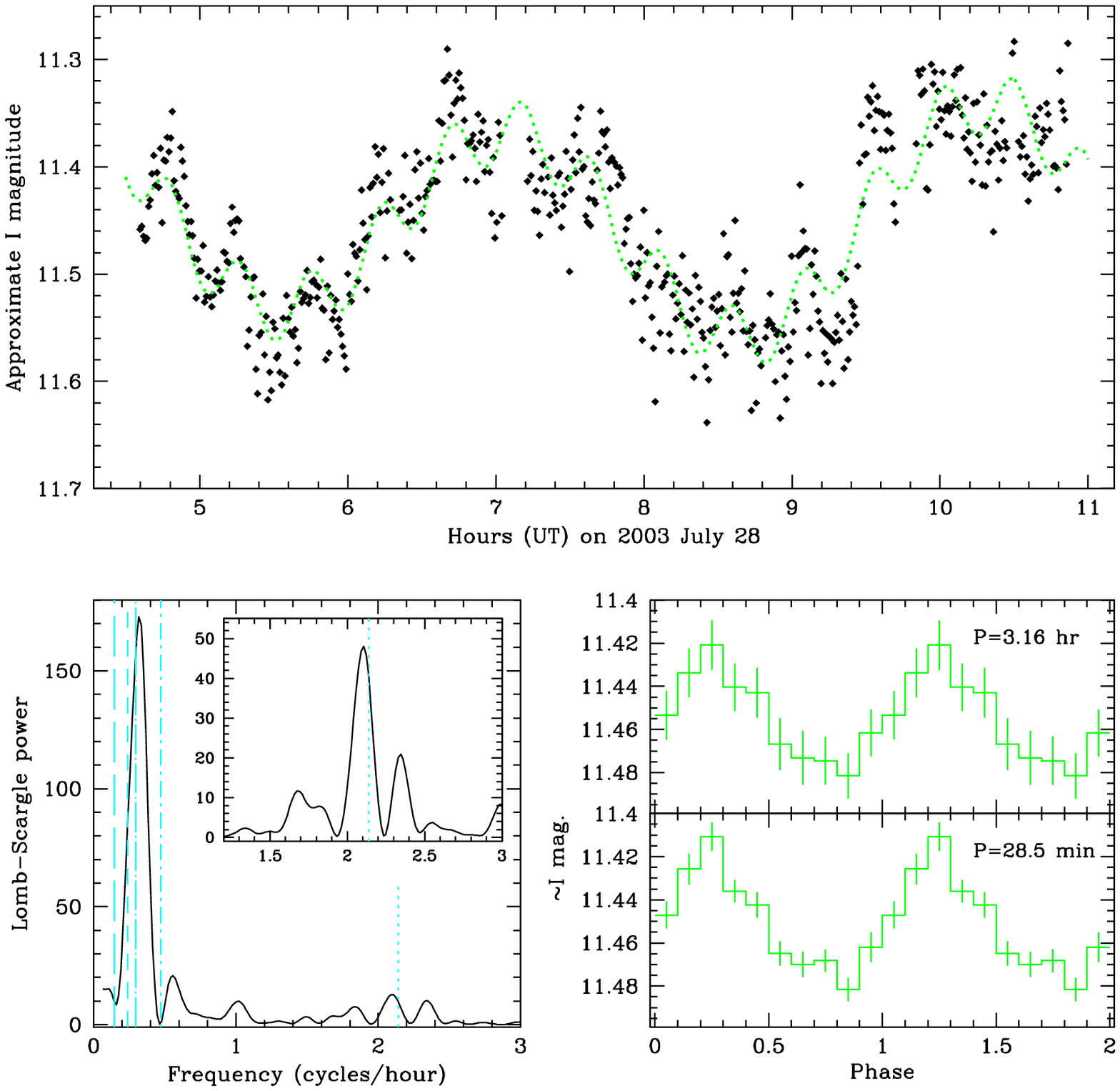}}}
\caption{Top panel: Differential $I$ band lightcurve of \targ, a triple-sinusoid fit with  $P=3.16, 6.32$~hrs (modelling the expected
  ellipsoidal variation) and 28.5 min is over-plotted. Lower left: Lomb-Scargle
  periodogram of $I$ data; the inset shows a close up in the vicinity of the 28.5 min peak, from data where the longer period sinusoids have
  been subtracted. A number of significant peaks  ($>99$\% confidence) are
  apparent, the largest now corresponding to the 3.1~hr modulation, while the rest are likely due to the prominent flickering.  The vertical
  bars indicate (left to right) the radial velocity determination of the binary orbital period (6.85~hrs), the X-ray signal (4.2~hrs), first
  harmonic of the orbit (3.42~hrs), and the first harmonic of the X-ray signal (2.1~hrs), plus the 28 min period from \ginga.  The large peak is reasonably consistent
  with half the orbital period, whilst the signal at 28.5 min is close to that found by \ginga.  However, we do note the presence of other peaks
  in that frequency range, indicating general flickering of which the 28.5 min signal happened to be the most significant. Lower right:  The $I$ band
  lightcurve folded on the periods shown and binned.\label{fig:mro_Ilc}}
\end{figure*}

Further quiescent lightcurves were obtained in 2003 July; the lightcurves, Lomb-Scargle periodograms and folded and
phase binned data are presented in Figures~\ref{fig:mro_Rlc} and ~\ref{fig:mro_Ilc}.  Once again fitting sinusoids, we find a modulation in the $R$ data at a period
consistent with the first harmonic of the 4.2~hr in the X-ray, $2.18\pm0.06$~hrs.  The power spectrum also shows a number of significant
peaks in the 20--50 min range, indicative of flickering.  In the $I$ band, the best fit period is $3.10\pm0.03$~hrs,
similar to half that of the orbit, as expected for an ellipsoidal modulation.  Indeed, adding a second sinusoid at double the period
does improve the fit.  However, in neither case are the periods formally consistent with those derived by \citet{hess88}.  On the other hand
the lightcurve does not quite span a full orbit, and together with the irregular/quasi-periodic flickering,  this may well account for the
discrepancy.  Lastly, removing the 3.1~hr sinusoid from the $I$ data we find that a significant modulation ($>99.0$\% confidence) remains
at $28.5\pm0.1$ min, which is in
agreement with the X-ray (from {\em Ginga}) and optical periods presented by \citet{szko90}.  Subdividing the lightcurve into four parts, we find
that the amplitudes are comparable at both the minima and maxima of the 3.10~hr cycle, although the frequency does shift, indicating that this
too is
most likely only quasi-periodic flickering.


\section{Evidence for a magnetic white dwarf} 
\subsection{Temporal}

The classic signature of an IP is the presence of two or more coherent periodic modulations corresponding to the binary orbit, white dwarf spin and possibly their beat.
In Table~\ref{tab:periods} we summarise all the periods reported for \targ\ in the literature and from our own (re)analyses.  The orbital period was definitively
measured by \citet{hess88}, whereas the existence of a second periodicity, the spin period, has been reported, revised and disputed over time
\citep[see][]{szko86,hell90,szko90,rose94}. The candidates (at \til0.5 and 1~hrs) are not, however, present in our long, uninterrupted \cxo\
dataset.  We have found such periodicities in consecutive nights of optical photometry, but neither the modulation nor period are stable,
the latter 
differing by \til20\% from one night to the next.  It seems most likely that these short periods are transient quasi-periodic behavior,
possibly fairly stable flickering.
\begin{deluxetable*}{lccccccc}
\tablewidth{0pt}
\tablecaption{Summary of Detected Periodicities in \targ\label{tab:periods}
}
\tablehead{
\colhead{Date} & \colhead{Observatory} & \colhead{State\tablenotemark{*}} &
\colhead{Wave band} &
\colhead{Period (hrs)} & \colhead{Percentage} & \colhead{Identification} & \colhead{References}\\
& (keV)&&&&\colhead{amplitude}&&}
\startdata
1980 June 15-17& Mt.Wilson& q& $H\beta$&         6.0 +\- 0.1 &      \nodata& $P_{orb}$&   2\\
1983 July 09-14 &{McDonald}&q& Abs. lines&$6.8475\pm0.0023$&\nodata&  $P_{orb}$ & 1\\
 &&&$H\gamma$,$H\beta$ &$6.820\pm0.076$&\nodata&  $P_{orb}$ & 6\\
 &&&Eq.W. &$3.272\pm0.009$&\nodata&  $P_{orb}/2$ (?)&6\\
1984 Sept 25 & {\em EXOSAT}/ME &q & X-ray(1.5--8.5) & $1.00\pm0.06$ & \til30& Flickering (?) & 2, 3\\
1988 Aug 8 &    SAAO\tablenotemark{a}  &    q& $H\beta$- V/R &   \til6.8&\nodata & $P_{orb}$&   3\\
 &&&                     Eq.W.&           \til6.8,\til1.5 &\nodata & $P_{orb}/4$ (?)&   3\\
1988 Sept 14--17 & {\em Ginga}&ss  & X-ray(2--6) & $2.105\pm0.005$\tablenotemark{b} & $26\pm4$& $P_{spin}/2$& 4\\
& & & & $4.210\pm0.010$\tablenotemark{b} & & $P_{spin}$& 4\\
& & & &  $0.47\pm0.02$ & \til10&Flickering (?)& 5 \\
1988 Sept 14--17 & KPNO &ss & Optical: V &$0.47$(fixed) & \til8&Flickering (?)& 5 \\
1989 July 06 & SAAO&q  & Optical: V & \til2.2 & \til20 & $P_{spin}/2$&3, 4\\
1991 Mar 28 & {\em ROSAT}&\nodata & X-ray (0.5--2.5) & $4.43\pm0.08$ & $92\pm8$& $P_{spin}$&4\\
2002 May 30, 31 & {\em Chandra}&q  & X-ray(0.3--10) & $4.23\pm0.05$ \tablenotemark{b}  &$65\pm5$ & $P_{spin}$&4\\
		&&			&&  $2.12\pm0.03$ \tablenotemark{b}  & & $P_{spin}/2$&4\\
2002 May 31 & BO &q & Optical: V &\til4.2 & \til20& $P_{spin}$& 4\\
2003 Aug 25 & MRO&q  & Optical: R & $2.18\pm0.06$ & $10\pm1$& $P_{spin}/2$& 4\\
2003 Aug 28 && & Optical: I & $3.16\pm0.02$ \tablenotemark{c} &$20\pm1$&  $P_{orb}/2$?& 4\\
&&&& $ 6.32\pm0.04$\tablenotemark{c} & & $P_{orb}$?& 4\\
&&&& $ 0.475\pm.002$ & $6\pm1$ & Flickering (?) & 4\\

\enddata
\tablenotetext{*}{q -- quiescence, ss -- stand-still} 
\tablenotetext{a}{South African Astronomical Observatory}
\tablenotetext{b}{Periods constrained to ratio of 2, i.e. fundamental plus first harmonic}

\tablenotetext{c}{Sinusoids constrained to be in-phase, with a period ratio of 2, in order to model the expected ellipsoidal modulation. Note
  the formal uncertainty quoted here is almost certainly an underestimate.}
\tablerefs{(1) \citet{hess88}. (2) \citet{szko86}.  (3) \citet{hell90}.   (4) This paper. (5) \citet{szko90}.  (6) Hessman (private communication)}
 \end{deluxetable*}

Instead, we have now found another, stronger candidate for the white dwarf spin period, at a surprisingly long period \til 4.2~hrs. A similar
periodicity (4.5~hrs) was first reported in the \rosat\
observation, though this dataset has rather sparse sampling. In our \cxo\ lightcurve we have found a prominent modulation at 4.23~hrs, with a probable first harmonic
at 2.12~hrs, essentially consistent with the \rosat\ result. Our reanalysis of the \ginga\ dataset has also uncovered the same period (to within
the larger \cxo\ \til 1\% uncertainties), with the
first harmonic dominating.  Hence in X-rays either the 4.2~hr or its first harmonic appear persistently in observations from 1988 to 2002.  Furthermore, the same periodicities are present in the optical (V and R bands), both in 2002
 (simultaneous with the \cxo\ observations) and 2003.  
From the simultaneous observation we found a $180^\circ$ phase offset between X-ray and optical modulations on the 4.23~hr period (note: a similar result
was evident in the 28 min modulation observed in simultaneous {\em Ginga} and
optical data, Szkody et al. 1990); this is
consistent with the the effects of a reprocessed beam. 
Unfortunately, none of the X-ray data have sufficient S/N to test conclusively whether the modulation exhibits the expected energy dependence, but 
neither do they exclude that possibility.  We should also note that the 4.2~hr signal is  not apparent in a reanalysis of the 1988 optical spectroscopy
(Hessman 2004, priv. comm.).  However, in at least one other case, EX Hya during outburst \citep{hell89b}, it appears that any signal on the
white dwarf spin, emitted from the magnetic disruption/corotation radius, is washed out by the line emission from the outer disk.  

While a white dwarf spin is the easiest explanation for the 4.2~hr period, the  physical origin of this long $P_{spin}$ is problematic.  First, if we assume that \targ\ possesses a fairly
typical field for an IP
(1--10 MG)  and mass transfer rate of $10^{-9}-10^{-10}$\msun yr$^{-1}$, then the Alfv\'{e}n radius is \til$2\times10^{10}$cm for spherical
accretion, or half that in the disk case. Depending on the actual values used this radius  could easily be smaller than both the radius of closest approach of the stream from the secondary and that of the initial ring that would form, hence
formation of a disk is certainly possible \citep{warn95}. However, if we also assume that the corotation radius is similar, one requires a
$P_{spin}\sim10$ min, while  $P_{spin}= 4.23$~hrs would place the corotation radius outside the Roche lobe ($R_{co}\geq9.0\times10^{10}$cm
as compared to the white dwarf-$L_1$ point separation $b\geq7.8\times10^{10}$cm).  Even taking  $P_{spin}=2.1$~hrs,
which is in itself difficult to reconcile with the data,
leads to a corotation radius comparable to the size of a typical disk; hence the entire disk would be non-Keplerian and in a transitional state to full
magnetic dominance.  In addition, for disk-less accretion the size of the magnetic moment and in turn surface field are unreasonable, even taking
$P_{spin}=2.1$~hrs, we estimate $B_1=850 (\dot{M}/\times10^{-9} \msun {\rm yr}^{-1})$ MG.

An alternate model was put forward by \citet{king99} to account for the spin-orbit equilibrium at $P_{spin}/P_{orb}=0.68$ of the short period
IP EX Hya. In this instance, the magnetic moment is such that the corotation radius is in fact comparable with the distance to the $L_1$
point. \citet{nort03} have more recently extended this work in an attempt to explain the entire range of spin-orbit equilibria
seen for IPs.  From their figure 2, we see that at $P_{orb}\equiv7$hrs, and  $P_{spin}/P_{orb}=0.6$, we are within the uncertainties of the
model (which does assume $M_1=0.7\msun$ in any case) in obtaining a suitable equilibrium.  But a longer period (larger) system like \targ\ requires
$\mu\sim10^{35}$Gcm$^{-3}$, or equivalently $B_1\sim300$ MG,
comparable to the most magnetic polars.

 An alternative non-magnetic origin for the 4.23~hr modulation could be stream overflow, which brightens a particular region of the disk. From
\citet{hess88} tables III and IV, the Keplerian period at the outer radius of the disk $r_J$ is $P_{Kep}(r_J)/P_{orb}=(1+q)^{1/2}(r_J/a)^{3/2}$=0.07, which is much
  too small.  Conversely, to obtain a ratio of 0.62, would require $q\sim0.007$, again highly implausible.

A final possibility is that we could be seeing \targ\ in an non-equilibrium state, where \.{M} was greatly reduced on Myr timescales compared
to what we now observe.  However, this is an ad hoc assumption which has no satisfactory explanation.

\subsection{Spectral}
It is instructive to compare our spectral results on \targ, with those of previous X-ray observations and \cxo\ results on other CVs.  Given its Z Cam classification, the mass transfer rate for \targ\ should be high.  Instead the value
\til$2\times10^{-11} M_{\odot}\; {\rm
	 yr}^{-1}$ places it in the middle of the five CVs that \citet{muka03} fit with cooling flows.  Moreover, we may compare this to an estimate of
the mass accretion rate from the X-ray luminosity, $L_X(2-10 {\rm keV})=8\times10^{33}\ergsec$, which yields \.{M}=$7\times10^{-12} M_{\odot}\; {\rm
	 yr}^{-1}$, assuming half of the energy is released as X-rays. But as \citet{pern03} comment in the case of WX
Hyi (where the cooling flow yields \.{M} a factor of five less than obtained from fits to the UV continuum), these X-ray fits give  \.{M} in the outer disc,
which may well not be accreting during quiescence, as the disk builds up for the next outburst. 

As previously noted, for a thermal bremsstrahlung fit ($N_H=0.7\times10^{22}$cm$^{-2}$, $kT_{Br}=32$~keV), the {\em
  EXOSAT} data yielded a high 2--10~keV flux
of $8-8.4\times10^{-11}\ergsqcmsec$, suggesting a magnetic system. For a thermal spectrum plus Fe line at 6.8~keV, the {\it
  Ginga} data gave a flux value a factor of two
lower, together with an even higher column ($1.6\times10^{22}$cm$^{-2}$), and a lower temperature ($kT_{Br}=14$~keV). However, at the time of
  the \ginga\ observation \targ\ 
was in a stand-still state, close to outburst, 
hence the lower flux and higher column can be expected. Our \cxo\ observation provides another look at the quiescent state,
but we have found a column and temperature intermediate between  {\em EXOSAT} and
{\it Ginga}, $1.0\times10^{22}$cm$^{-2}$ and \til20~keV respectively. Moreover, the flux given by the best cooling flow fit, \til$3\times10^{-11}\ergsqcmsec$ (2--10~keV),
is much lower than either of the past satellite observations. Thus, it appears
difficult to directly compare data taken at different outburst states and with
different instrument sensitivities.

If \targ\ contains a magnetic white dwarf, the X-rays should arise from a cooling region behind
the accretion shock, and a
shock model for the emission lines is appropriate.
\citet{muka03} found that polar spectra were better
fit by models of photoionized plasma.  Both polars and
intermediate polars must have emission both from the
cooling zone behind the accretion shock and from the
photoionized gas upstream.  The difference may be that
polars have much narrower accretion columns and higher
densities, and the recombination radiation from the
photoionized accretion stream scales as the density
squared.  The emission per unit accreted mass from
the shock, on the other hand, is independent of density.
Ratios of lines from 3s and 3d levels of iron ions
are a good diagnostic for photoionized plasmas
\citep{lied84}, and the ratios of the Fe XVII
lines and ratios of Fe XXIV lines confirm that the
emission from V426 Oph originates in a collisionally
ionized plasma.
 
The match between observed fluxes and the shock model
is far from perfect, but it is close enough to indicate
that the shock structure is basically correct.  In particular,
thermal conduction would tend to produce overly strong
O VIII and Fe XVII lines unless it is suppressed by
magnetic fields or turbulence.  It is probably not
advisable to make too much of the differences between
the shock model and the observations at present, because
accretion shocks are thermally unstable \citep{lang82,imam84}.  The
instability and secondary shocks can modify the emission
line spectrum at the factor of 2 level \citep{inne92}.


\section{Conclusions}
Our analysis of a long uninterrupted \cxo\ X-ray lightcurve of \targ\ together with optical photometry and reanalysis of archival X-ray data, has
revealed a persistent period at 4.2~hrs, unrelated to the known orbit.  In contrast, we find that previous candidate spin periods seen in past
X-ray studies at \til0.5 and 1~hrs are most likely non-persistent QPOs. Our simultaneous \linebreak[4] X-ray/optical photometry shows that the modulations are anti-correlated and therefore the
variations must arise from different parts of the
system; possibly due to the reprocessing of beamed X-radiation.  Indeed, the period of 4.2~hrs
is easiest to interpret as the rotation of the white dwarf, implying
that \targ\ is, after all, an IP, and making it the first long period IP with a similar spin-orbit equilibrium to EX Hya,
$P_{spin}/P_{orb}\sim0.6$.  However, this interpretation leads to problems with the magnetic field
strengths and mass accretion rates that are typical for intermediate polars.

The high-resolution spectrum  places \targ\ as a member of the group of CVs, including U Gem and EX Hya, where the X-rays are emitted from a
cooling flow. Indeed, more detailed modeling of the line fluxes for such a shock model lends further support to the magnetic CV interpretation.

\acknowledgments
With the death of Bob Fried on November 13th, 2003 the CV community has lost an active and dedicated observer.  We gratefully acknowledge his
years of collaboration. 

We are grateful to the referee, Rick Hessman, for his helpful review. Support for this work was provided by NASA through {\it Chandra} award
GO2-3033X. This research has made use of data obtained through the High Energy Astrophysics Science Archive Research Center Online Service, provided by the NASA/Goddard Space Flight Center.



\end{document}